\documentstyle[amstex,epsfig,color,graphicx,titlepage,12pt]{article}

\begin{document}

\renewcommand{\thesection}{\Roman{section}.} \baselineskip=24pt plus 1pt
minus 1pt

\begin{titlepage}
\vspace*{0.5cm}
\begin{center}

\LARGE\bf Comment on \textquoteleft Entanglement of photon-added nonlinear coherent states via beam splitter\textquoteright
\\[1.5cm]
\normalsize\bf Hadia Bint Monir and Shahid Iqbal

\end{center}


\vspace{7pt}

\begin{center}

Department of Physics, School of Natural Sciences,\\
National University of Sciences and Technology\\
Islamabad, Pakistan
\end{center}




\vspace{0.3cm}

\normalsize Beam splitter mediated entanglement of nonlinear photon-added coherent states was studied by Honarasa, Bagheri and Gharaati \cite{rmp}. We believe that this work contains errors in numerical computations of linear entropy and Mandel parameter which lead to wrong conclusions. We point out the possible errors and describe the necessary corrections. The conclusions based on our arguments are in complete agreement with previously published literature.

%


\vspace{0.2cm}

\end{titlepage}

\newpage

Entanglement is regarded as a key resource for quantum information processing and therefore, the generation of entangled states received a lot of attention. 
Recently, Honarasa, Bagheri and Gharaati \cite{rmp} investigated the entanglement generation at output modes of a beam splitter with a nonlinear photon-added coherent state and a vacuum state injected at the inputs. In particular, two types of nonlinear photon-added coherent states, namely Panson-Solomon coherent states and coherent states associated with P\"{o}schl-Teller potential, were used as one input of the beam splitter. 

Earlier, it has been pointed out that beam splitter mediated entanglement generation requires at least one input mode as nonclassical state \cite{entangle1}. In \cite{rmp}, the nonclassical character of the nonlinear photon-added coherent states, as an input state of the beam splitter, was analyzed by Mandel parameter (sub-Poissonian statics: an indicator of nonclassicality) and the entanglement generation at output states was quantified by linear entropy. Here we argue that the results presented by the plots of Mandel parameter and linear entropy concerning Panson-Solomon coherent states (displayed by Figures (3) and (1) of \cite{rmp}) are not correct and lead to wrong conclusions. 
%
%

In order to discuss the above mentioned results of \cite{rmp}, we rewrite the expressions involved in the computation of linear entropy and Mandel parameter.
The general form of photon-added nonlinear coherent states (introduced in Eq.~(2) of \cite{rmp}) is given as, 
\begin{equation}
\vert \alpha, f,k\rangle =N_{\alpha,f,k} \sum_{n=0}^{\infty}\frac{\alpha^{n}[f(n+k)]!\sqrt{(n+k)!}}{n![f(n)]!}\vert n+k\rangle,
\end{equation}
where, 
\begin{equation}
N_{\alpha,f,k} =\left(\sum_{n=0}^{\infty}\frac{\vert \alpha\vert^{2n}[f^{2}(n+k)]!(n+k)!}{(n!)^{2}[f^{2}(n)]!}\right)^{-1/2}
\end{equation}
is the normalization constant and $[f(n)]!:=f(n)f(n-1).......f(1)$. Here we will only discuss the case of Panson-Solomon coherent states for which $f(n)=q^{1-n}$ with $0\leq q \leq 1$ as given by Eq.~(13) of  \cite{rmp}.

The Mandel parameter $Q$ is defined as \cite{man},
\begin{equation}
Q=\frac{\langle n^2\rangle-\langle n\rangle^2}{\langle n\rangle}-1=\frac{\sigma^{2}}{\left\langle n\right\rangle }-1. \label{q}
\end{equation} 
The negative values of the Mandel parameter $-1\le Q<0$ (sub-Poissonian distribution) indicate the nonclassicality of arbitrary quantum states, such that, $Q=-1$ belongs to highly nonclassical Fock states and the limit $Q\rightarrow 0$ belongs to classical-like coherent states. Moreover, the linear entropy is defined \cite{lin} as
\begin{equation}
S=1-Tr(\rho_{a}^{2}), \label{s}
\end{equation} 
where $\rho_{a}$ is the reduced density matrix of a bipartite system under consideration, such as, that of output modes of beam splitter in the present case given by Eq.~(9) of \cite{rmp}. The numerical value of linear entropy ranges between zero (no entanglement) and one (maximum entanglement).
\begin{figure}
\centering
\includegraphics[width=0.45\textwidth]{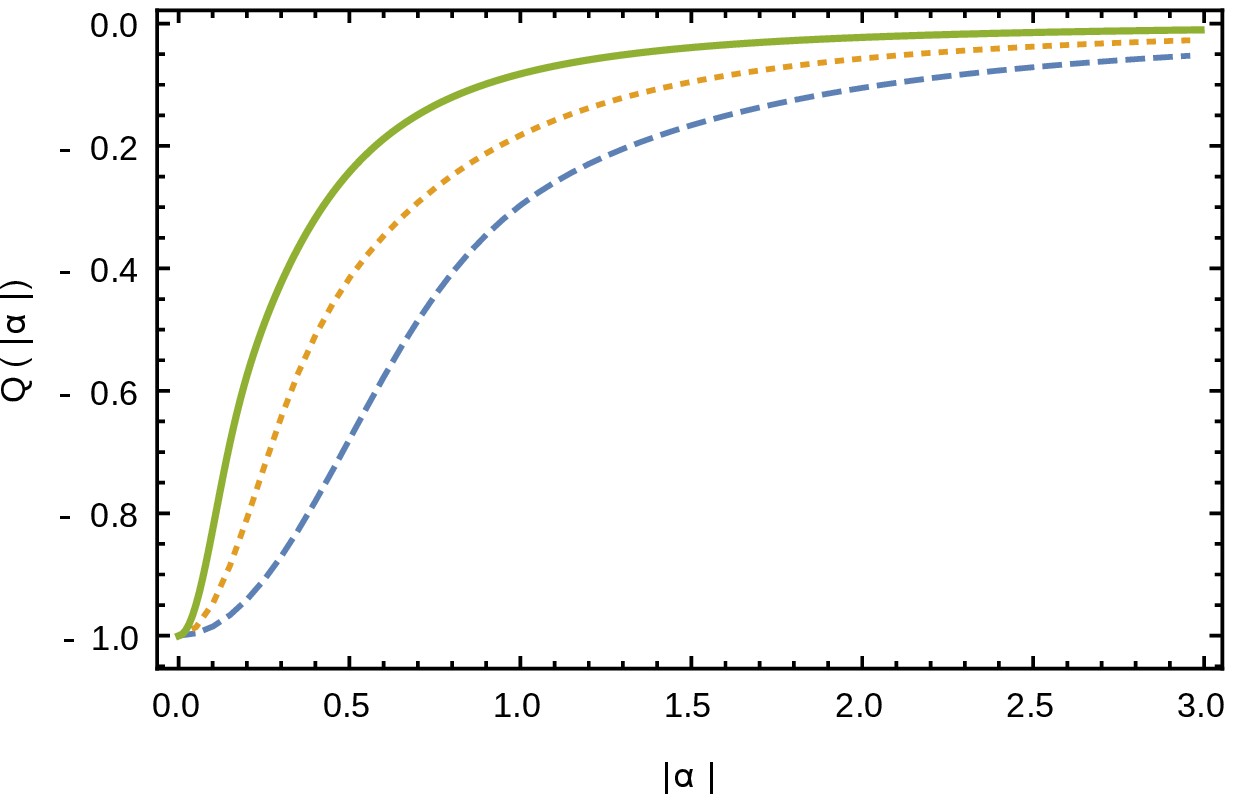}
\includegraphics[width=0.45\textwidth]{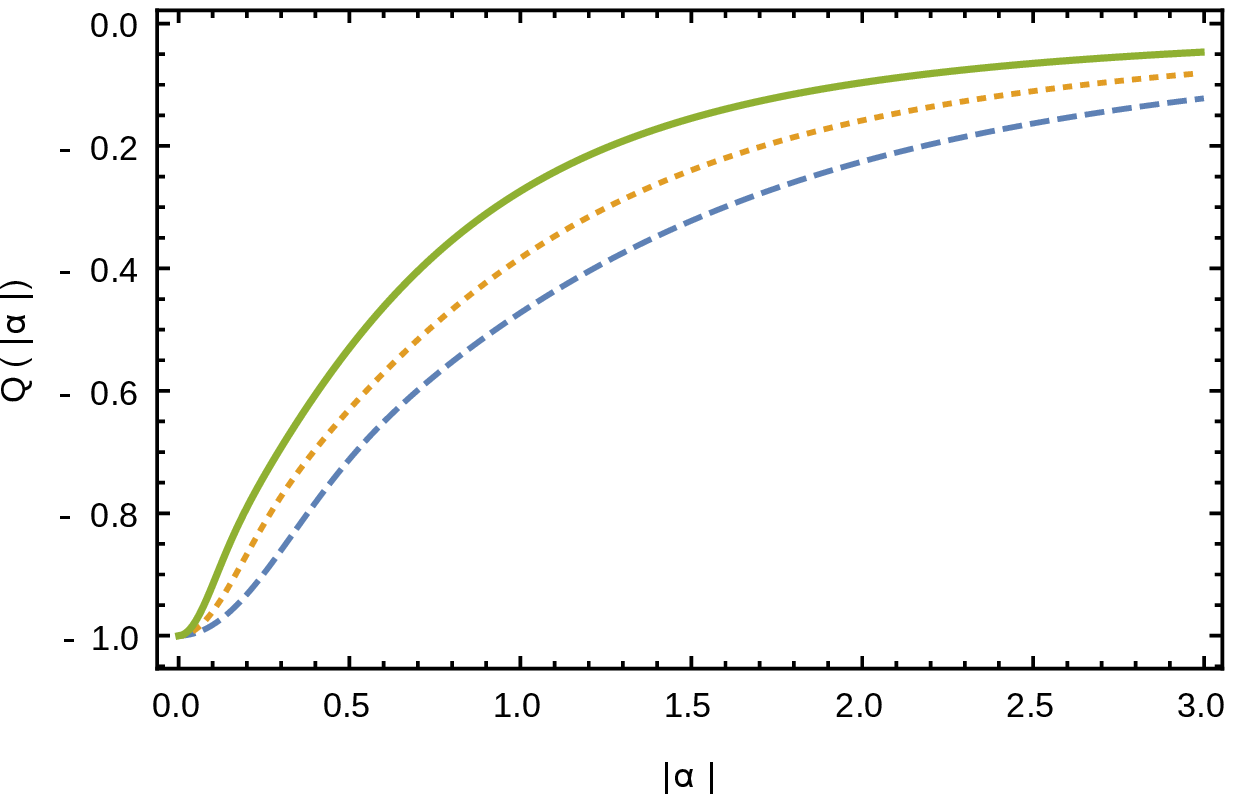}
\caption{The Mandel parameter $Q$ of photon-added nonlinear Panson-Solomon coherent states versus $\vert \alpha \vert$ for: (left) nonlinearity parameter $q=0.5$, $k=1$ (dashed line),
$k=2$ (dotted line), $k=3$ (solid line); (right) nonlinearity parameter $q=0.8$, $k=4$ (dashed line),
$k=6$ (dotted line), $k=8$ (solid line).} \label{mand}
\end{figure}

In Fig.~(1) of \cite{rmp},  the linear entropy $S$ was plotted for photon-added Panson-Solomon nonlinear coherent states as a function of the coherent state parameter $\vert \alpha\vert$.
Those plots indicate that the entanglement between the output states of beam splitter is maximum ($S=1$) for $\vert \alpha \vert=0$ which decreases sharply as the $\vert \alpha \vert$ increases. After reaching a minimum value at some particular value of $\vert \alpha \vert$ (which depends on number of photon added $k$), it increases sharply and attains the maximum value ($S=1$) again which stays the same for all higher values of $\vert \alpha \vert$.

In order to support their results (displayed by Fig. (1)) concerning entanglement generation, the authors of \cite{rmp} computed Mandel parameter $Q$ as a function of coherent state parameter $\vert \alpha\vert$ and the results was displayed in Figure (3). The plots in the Figure (3) indicate that the Panson-Solomon  photon-added nonlinear coherent states exhibit maximal nonclassicality ($Q=-1$) at $\vert \alpha \vert=0$ which decreases gradually as the $\vert \alpha \vert$ increases. However, after reaching its minimum value after short range of $\vert \alpha \vert$ (which depends on the photon excitation number $k$) the nonclassicality increases sharply to reach its maximal value ($Q=-1$) again which remains so for all higher values of $\vert \alpha \vert$.

Here we present our analysis on the results displayed in Figures (1) and (3) of \cite{rmp}. It is known fact \cite{entangle2,entangle3} that a $k$-photon-added coherent state $\vert \alpha, k\rangle$ approaches to the Fock state $\vert k\rangle $ (highly nonclassical with $Q=-1$) as $\vert \alpha\vert\rightarrow 0$. As the value of $\vert \alpha\vert$ increases from zero, the nonclassicality decrease from its maximal value. As a result, the entanglement (a nonclassical correlation) at output modes of the beam splitter should decrease when the nonclassicality of the inputs decreases \cite{entangle1}. Therefore, the results depicted in Figures (1) and (3) of \cite{rmp} seem consistent with the literature \cite{entangle1,entangle2,entangle3} only for lower of $\vert \alpha\vert$. However, after a short range of $\vert\alpha\vert$ (which seems depending on $k$), the deviation of the nonclassicality (depicted in Figure (3)) and entanglement (depicted in Figure (1)) towards higher values is very unexpected. We believe that this deviation is a consequence of computational errors, which we prove numerically in the following.


For our quantitative analysis we compute the Mandel parameter $Q$ for the case of Panson-Solomon nonlinear coherent states. For convenience, we rewrite the Eqs. (17) and (18) of \cite{rmp} as
\begin{eqnarray}
\langle n\rangle&=& N_{\alpha,f,k}^{2}\sum_{n=0}^{\infty}\frac{\vert \alpha\vert^{2n}[f^{2}(n+k)]!(n+k)!}{(n!)^{2}[f^{2}(n)]!}(n+k), \label{n1}\\ 
\langle n^{2}\rangle&=&N_{\alpha,f,k}^{2}\sum_{n=0}^{\infty}\frac{\vert \alpha\vert^{2n}[f^{2}(n+k)]!(n+k)!}{(n!)^{2}[f^{2}(n)]!}(n+k)^{2}.\label{n2}
\end{eqnarray}
We use Eqs. (\ref{n1}) and (\ref{n2}) into Eq.~(\ref{q}) to numerically compute Mandel parameter $Q$. In Figure (1) of this article, we plot the Mandel parameter $Q$ as a function of $\vert \alpha\vert$ for different values of nonlinearity parameter $q$ and number of added-photons $k$. It is obvious from these plots that the Mandel parameter $Q$ increases gradually from its minimal value as $\vert \alpha\vert$ increases. In contrast to the results displayed in Figures (3) of  [1] there is no sudden deviation from the usual trend at higher values of $\vert \alpha\vert$. Hence the nonclassicality of photon added nonlinear Panson-Solomon coherent states decreases as the parameter $\vert \alpha\vert$ increases which is in complete agreement with earlier published results \cite{entangle2, entangle3}. Moreover, since the nonclassicality of the input state is a prerequisite for the beam splitter mediated entanglement \cite{entangle1}, therefore the results of the Figure (1) of this article suggest that the entanglement at output modes should also decrease as $\vert \alpha\vert$ increases. This leads us to conclude that the results depicted in Figure (1) of \cite{rmp} are also incorrect.

As a matter of fact, we probe the possible error in the numerical computation of the linear entropy and Mandel parameter, depicted in Figures (1) and (3) of \cite{rmp},
that led to the wrong results. Based on the early studies \cite{c1,c2,c3,c4}, we believe that the authors of \cite{rmp} did not take the required accuracy into account in the computation of the linear entropy and the Mandel parameter of photon-added Panson-Solomon nonlinear coherent states. In fact, in the numerical computation of an expression involving a summation over infinite terms, one has to truncate the series sum after a finite number of terms ($n_{max}$). The maximum number of terms, $n_{max}$, is chosen such that, the required accuracy is achieved. If the $n_{max}$ is not chosen properly, the computation leads to wrong results (for more details, see \cite{c1,c2}). To demonstrate explicitly the effect of wrong choice of $n_{max}$, we compute the Mandel parameter of photon-added Panson-Solomon nonlinear coherent states by choosing the values of $n_{max}$ arbitrarily and the results are displayed in Figure (2) of this article. It is obvious from this Figure that an arbitrary choice of $n_{max}$ leads to the similar erroneous results as have been depicted in Figure (3) of \cite{rmp}. 

%
%
%
\begin{figure}
\centering
\includegraphics[width=0.8\textwidth]{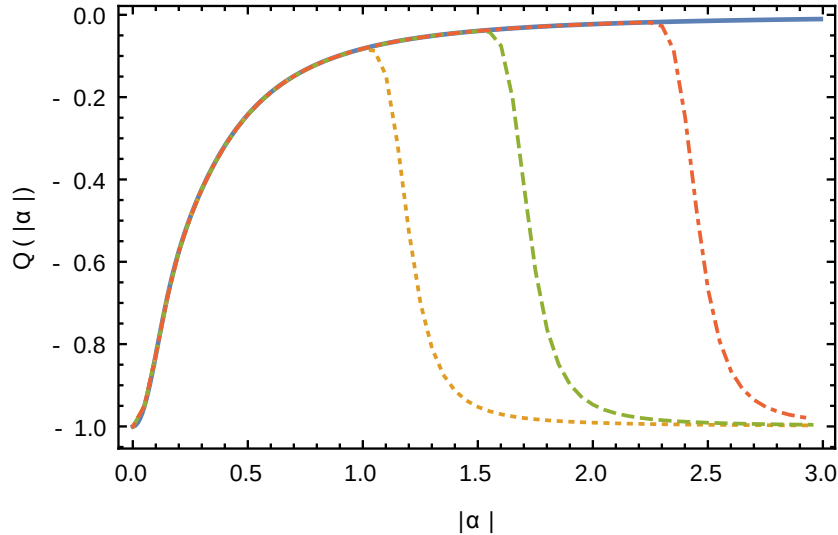}
\caption{The Mandel parameter $Q$ of photon-added nonlinear Panson-Solomon coherent states ($k=3$) versus $\vert \alpha \vert$ for $n_{max}=100$ (dotted line),
$n_{max}=200$ (dashed line), $n_{max}=400$ (dot-dashed line),
$n_{max}=700$ (solid line).} \label{mom}
\end{figure}

Based on the above arguments we conclude that the results concerning linear entropy and Mandel parameter of photon added Panson-Solomon nonlinear coherent states presented in \cite{rmp} are incorrect. In contrast to the results of \cite{rmp}, it is stated that the  nonclassicality of the photon-added nonlinear coherent states and the entanglement generation between output modes of the beam splitter are decreasing as the coherent state parameter $\vert \alpha\vert$ increases.

The present work is supported partially by The World Academy of Sciences (TWAS) through grant No. 15-306 RG/ITC/AS$\_$C -- FR3240288940.

\end{document}